\documentclass[12pt]{article}
\usepackage{graphicx,amsfonts,amsmath,amssymb,mathrsfs,url}
\usepackage[usenames,dvipsnames]{xcolor}

\title{On Bohmian Mechanics, Particle Creation, and Relativistic Space-Time: Happy 100th Birthday, David Bohm!}
\author{
Roderich Tumulka\footnote{Mathematisches Institut,
     Eberhard-Karls-Universit\"at, Auf der Morgenstelle 10, 72076
     T\"ubingen, Germany. E-mail:
     roderich.tumulka@uni-tuebingen.de}
}
\date{June 1, 2018}

\newcommand{\Hilbert}{\mathscr{H}}
\newcommand{\foliation}{\mathscr{F}}
\newcommand{\Fock}{\mathscr{F}}
\newcommand{\conf}{\mathcal{Q}}
\newcommand{\Q}{\mathcal{Q}}

\newcommand{\sS}{\mathscr{S}}

\renewcommand{\Im}{\mathrm{Im}}
\newcommand{\RRR}{\mathbb{R}}
\newcommand{\CCC}{\mathbb{C}}
\newcommand{\SSS}{\mathbb{S}}
\newcommand{\scp}[2]{\langle #1|#2 \rangle}
\newcommand{\pr}[1]{| #1 \rangle \langle #1 |}

\newcommand{\orig}{{\mathrm{orig}}}
\newcommand{\cutoff}{{\mathrm{cutoff}}}
\newcommand{\vk}{\boldsymbol{k}}
\newcommand{\vx}{\boldsymbol{x}}
\newcommand{\vy}{\boldsymbol{y}}

\newcommand{\vQ}{\boldsymbol{Q}}
\newcommand{\vomega}{\boldsymbol{\omega}}
\newcommand{\vzero}{\boldsymbol{0}}

\newcommand{\domain}{\mathscr{D}}
\newcommand{\be}{\begin{equation}}
\newcommand{\ee}{\end{equation}}

\begin{document}
\maketitle
\begin{abstract}
The biggest and most lasting among David Bohm's (1917--1992) many achievements is to have proposed a picture of reality that explains the empirical rules of quantum mechanics. This picture, known as pilot wave theory or Bohmian mechanics among other names, is still the simplest and most convincing explanation available. According to this theory, electrons are point particles in the literal sense and move along trajectories governed by Bohm's equation of motion. In this paper, I describe some more recent developments and extensions of Bohmian mechanics, concerning in particular relativistic space-time and particle creation and annihilation.

\medskip

  \noindent 

  Key words: de Broglie-Bohm interpretation of quantum mechanics; pilot wave; interior-boundary condition; ultraviolet divergence; quantum field theory. 
\end{abstract}

\section{Introduction}

In 1952, David Bohm \cite{Bohm52} solved the biggest of all problems in quantum mechanics, which is to provide an explanation of quantum mechanics. (For discussion of this problem see, e.g., \cite{Bri16,Nor2018,einstein,Bell90}.) His theory is known as Bohmian mechanics, pilot-wave theory, de Broglie--Bohm theory, or the ontological interpretation. This theory makes a proposal for how our world might work that agrees with all empirical observations of quantum mechanics. Unfortunately, it is widely under-appreciated. It achieves something that was often (before and even after 1952) claimed impossible: To explain the rules of quantum mechanics through a coherent picture of microscopic reality.

\begin{figure}[h]
\begin{center} 
  \includegraphics[width=0.8\textwidth]{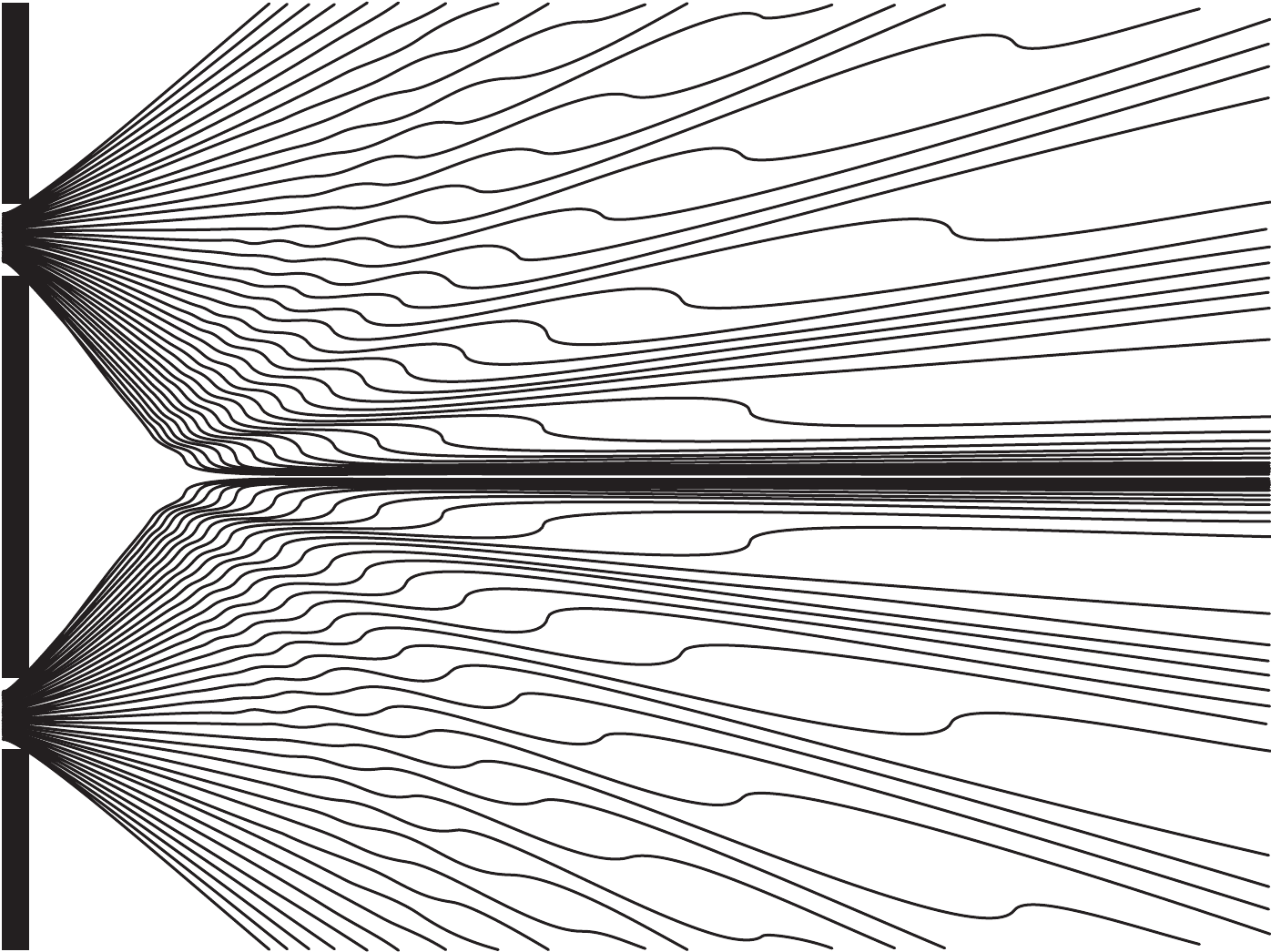}
\end{center}
 \caption{Several possible trajectories for a Bohmian particle in a double-slit setup, coming from the left. (Reprinted from \cite{DT}, based on a figure in \cite{philippidis79}.)}\label{pic}
\end{figure}

In the following, I will briefly review Bohmian mechanics and then discuss some extensions of it that were developed in recent years. For textbook-length introductions to Bohmian mechanics, see \cite{BH,DT,Bri16,Nor2018}; for a recent overview article, see \cite{Tum18}.

\subsection{Significance of Bohmian Mechanics}

Bohmian mechanics is remarkably simple and elegant. In my humble opinion, some extension of it is probably the true theory of quantum reality. Compared to Bohmian mechanics, orthodox quantum mechanics appears rather incoherent. In fact, orthodox quantum mechanics appears like the narrative of a dream whose logic does not make sense any more once you are awake although it seemed completely natural while you were dreaming. (E.g., \cite{einstein,Bri16}.) 

According to Bohmian mechanics, electrons and other elementary particles are particles in the literal sense, i.e., they have a well-defined position $\vQ_j(t)\in \RRR^3$ at all times $t$. They have trajectories. These trajectories are governed by Bohm's equation of motion (see below). In view of the widespread claim that it was impossible to explain quantum mechanics, it seems remarkable that something as simple as particle trajectories does the job. So what went wrong in orthodox QM? Some variables were left out of consideration: the particle positions!

\subsection{Laws of Bohmian mechanics}

According to non-relativistic Bohmian mechanics of $N$ particles, the position $\vQ_j(t)$ of particle $j$ in Euclidean 3-space moves according to Bohm's equation of motion,
   \be\label{Bohm}
   \frac{d\vQ_j}{dt} = \frac{\hbar}{m_j} \text{Im}\, \frac{\psi^* \nabla_j \psi}{\psi^* \psi}(\vQ_1,\ldots, \vQ_N)
   \ee
for every $j=1,\ldots,N$. If some particles have spin, then $\psi^* \phi$ means the inner product in spin space. The wave function $\psi$ of the universe evolves according to the Schr\"odinger equation,
  \be\label{Schr}
  i\hbar\frac{\partial \psi}{\partial t} = -\sum_j \frac{\hbar^2}{2m_j} \nabla_j^2 \psi + V\psi \,.
  \ee
The initial configuration $Q(0) = (\vQ_1(0), \ldots, \vQ_N(0))$ of the universe is random with probability density\footnote{Actually, the point $Q(0)$ need not be truly random; it suffices that $Q(0)$ ``looks typical'' with respect to the statistical properties of the ensuing history $t\mapsto Q(t)$ \cite{DGZ92}, much like the number $\pi$ is not truly random but its decimal expansion looks like a typical sequence of digits.}
  \be\label{Born}
  \rho = |\psi_0|^2 \,.
  \ee

\subsection{Properties of Bohmian Mechanics}

It follows from \eqref{Bohm}--\eqref{Born} that at any time $t\in\RRR$, $Q(t)$ is random with density $\rho_t = |\psi_t|^2$ (``equi\-variance theorem'' or ``preservation of $|\psi|^2$''). It follows further, by a theorem akin to the law of large numbers, that subsystems of the universe with wave function $\varphi$ will always have configurations that look random with $|\varphi|^2$ distribution \cite{DGZ92}. This fact, known as ``quantum equilibrium,'' is the root of the agreement between the empirical predictions of Bohmian mechanics and the rules of the quantum formalism.

For an example of equivariance and quantum equilibrium, Figure~\ref{pic} shows a selection of trajectories for the double-slit experiment with roughly a $|\varphi|^2$ distribution, where $\varphi$ is a 1-particle wave function. The equivariance theorem implies that the arrival places on the right (where one may put a screen) are $|\varphi|^2$ distributed; thus, more particles arrive where $|\varphi|^2$ is larger. John Bell commented \cite{Bell86b}:
\begin{quotation}
``This idea seems
to me so natural and simple, to resolve the wave--particle dilemma in
such a clear and ordinary way, that it is a great mystery to me that
it was so generally ignored.''
\end{quotation}

Bohmian mechanics is clearly non-local (i.e., involves faster-than-light influences) because, according to \eqref{Bohm}, the velocity of particle $j$ depends on the simultaneous positions of all other particles $\vQ_1,\ldots,\vQ_N$. Of course, Bell's theorem \cite{Bell87b} shows that every theory in agreement with the empirical facts of quantum mechanics must be non-local.

Bohmian mechanics avoids the problematical idea that the world consists only of wave function.
It provides precision, clarity, and a clear ontology in space-time.
And it allows for an analysis of quantum measurements, thus replacing the postulates of orthodox quantum mechanics by theorems.

\section{Extension of Bohmian Mechanics to Particle Creation}
\label{sec:crea}

\begin{figure}[h]
\begin{center} 
\includegraphics[width=0.66\textwidth]{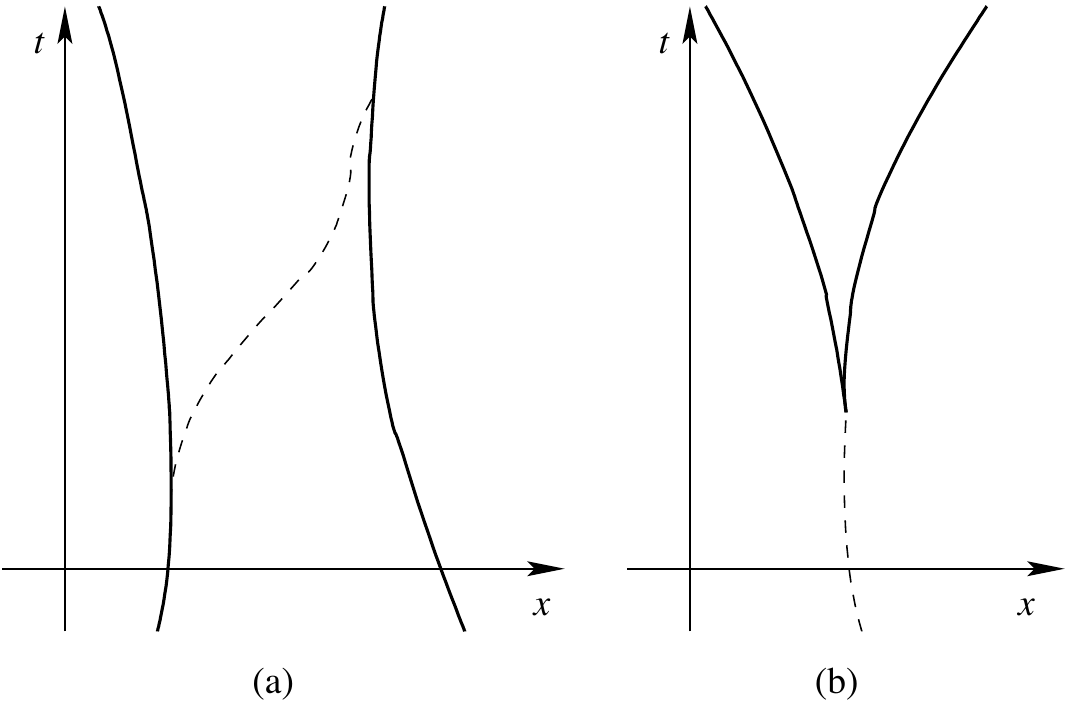}\\[5mm]
\end{center}
 \caption{Possible patterns of particle world lines in theories with particle creation and annihilation. (a) A boson (dashed world line) is emitted by a fermion and absorbed by another. (b) A boson (dashed world line) decays into two fermions. (Reprinted from \cite{crlet}.)}\label{crletf1}
\end{figure}

Bohmian mechanics has been successfully extended so as to incorporate particle creation. 
In theories with particle trajectories, particle creation and annihilation mean that trajectories can begin and end (Figure~\ref{crletf1}). Perhaps the most plausible picture would have them begin and end on the trajectories of other particles.

Particle creation and annihilation come up particularly in quantum field theory (QFT); since we want to connect them with particle trajectories, we will make use of the particle-position representation of QFTs, a representation used also independently of the Bohmian approach, for example in  \cite{lp:1930,schweber:1961,Nel64}. The state vector then is a vector in Fock space $\Fock$,
\be
\psi\in \Fock = \bigoplus\limits_{n=0}^\infty \Hilbert_n\,,
\ee
or perhaps in the tensor product of several Fock spaces. Here, the $n$-particle Hilbert space $\Hilbert_n$ (also called the $n$-particle sector or simply $n$-sector of $\Fock$) is the symmetrized or anti-symmetrized $n$-th tensor power of the 1-particle Hilbert space $\Hilbert_1$. The position representation of $\psi\in\Fock$ is a function on the configuration space of a variable number of particles,
\be\label{confdef}
\Q = \bigcup\limits_{n=0}^\infty \RRR^{3n} \,,
\ee
and $|\psi|^2$ defines a probability distribution on $\Q$. Here, $\RRR^{3n}$ is called the $n$-sector of $\Q$. 
(In fact, it is often desirable to use \emph{unordered} configurations $\{\vx_1,\ldots,\vx_N\}$ because in nature, configurations are not ordered. In \eqref{confdef} and in the following, we use \emph{ordered} configurations $(\vx_1,\ldots,\vx_N)$ because that allows for easier notation.)

\subsection{Bell's Jump Process (in Its Continuum Version)}

\begin{figure}[h]
\begin{center} 
\includegraphics[width=0.6\textwidth]{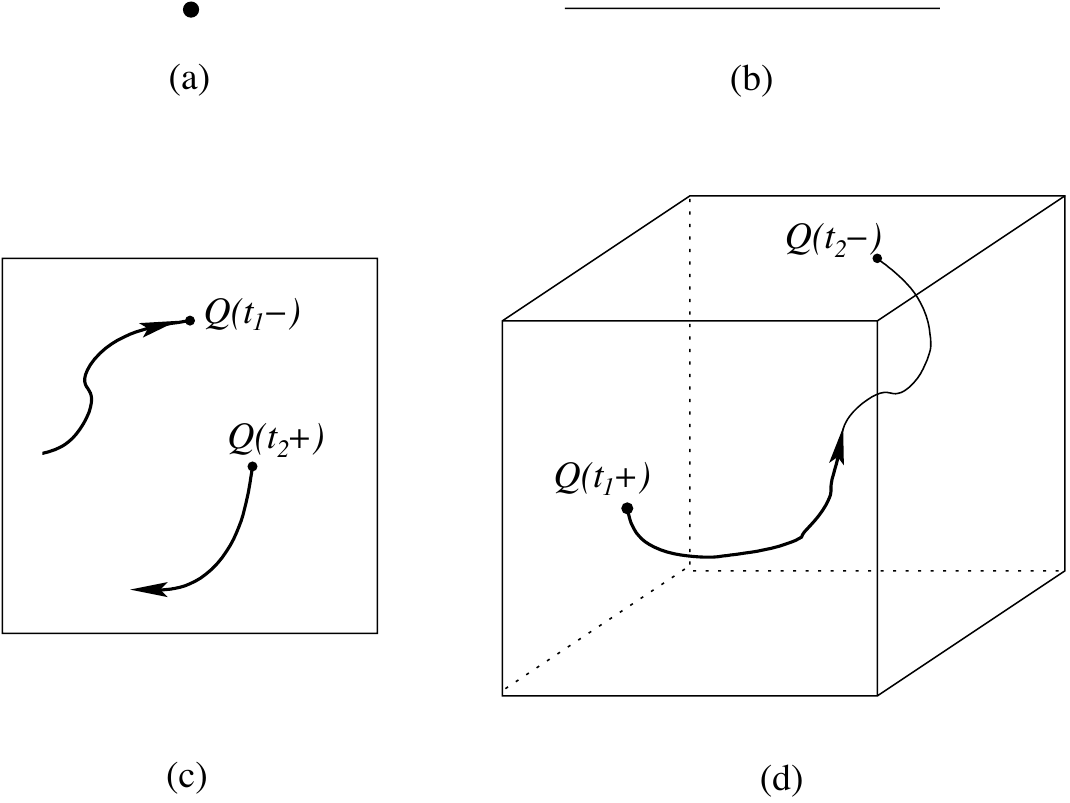}
\end{center}
 \caption{The configuration space \eqref{confdef} of a variable number of particles; drawn are, for space dimension $d=1$, the first four sectors: (a) The 0-particle sector has a single element, the empty configuration; (b) the 1-particle sector is a copy of physical space; (c) the 2-particle sector; (d) the 3-particle sector. In addition, the configuration curve corresponding to Figure~\ref{crletf1}(a) is drawn; it jumps at time $t_1$ from the 2-particle sector to the 3-particle sector and at time $t_2$ back. (Reprinted from \cite{crlet}.)}\label{crletf2}
\end{figure}

Here is the natural extension of Bohmian mechanics to particle creation \cite{Bell86,DGTZ03,crlet,crea2B,Vi93,Vi17}; Bell \cite{Bell86} considered this on a lattice, but it can be set up as well in the continuum \cite{DGTZ03,crlet,crea2B}, and we will directly consider this case. The configuration curve $Q(t)$ will jump one sector up (respectively, down) whenever a particle is created (respectively, annihilated), see Figure~\ref{crletf2}. 

According to (the continuum version of) Bell's proposal, jumps (e.g., from the $n$-sector to the $n+1$-sector) occur in a \emph{stochastic} way, with rates governed by a further law of the theory. This means that, according to this theory, jumps occur spontaneously as an element of irreducible randomness in nature; they are not pre-determined by any further variables (``hidden'' or not). It was not the point of Bohmian mechanics to restore determinism but to hypothesize what actually happens in the microscopic reality; if the most convincing hypothesis turns out to be deterministic (as it does for fixed particle number), then that is fine, if not, that is fine, too. Here, the randomness in the jumps is relevant to ensuring that after particle creation, the configuration is still $|\psi|^2$ distributed.

Mathematically, $(Q(t))_{t\in\RRR}$ forms a stochastic process, in fact a Markov jump process. Between jumps, Bohm's equation of motion applies. The law governing the jumps reads as follows:  Given that the present configuration $Q(t)$ is $q'\in\Q$, the rate (i.e., probability per time) of jumping to a volume element $dq$ around $q\in\Q$ is
\be\label{Belljumprate}
  \sigma^{\psi}(q'\to dq) = 
  \frac{\max\bigl\{0,\tfrac{2}{\hbar} \, \Im \, 
  \scp{\psi}{q}\scp{q}{H_I|q'}\scp{q'}{\psi} \bigr\} }
  {\scp{\psi}{q'}\scp{q'}{\psi}} dq \,.
\ee
Here, $H_I$ is the interaction Hamiltonian as in $H=H_0+H_I$ with $H_0$ the free Hamiltonian. More generally, $\pr{q}\, dq$ could be replaced by a PVM (projection-valued measure) or a POVM (positive-operator-valued measure) $P(dq)$ on $\Q$ (and $\pr{q'}$ by $P(dq')$, as factors of $dq'$ would cancel out). 
Since $H_I$ usually links only to the next higher and lower sector, only jumps to the next higher or lower sector are allowed by \eqref{Belljumprate}. 

The jump rate \eqref{Belljumprate} is so designed as to entail an equivariance theorem \cite{crea2B}: that is, if $Q(0)$ is $|\psi_0|^2$ distributed (that is, abstractly speaking, $\scp{\psi_0}{P(\cdot)|\psi_0}$ distributed), then at every $t\in\RRR$, $Q(t)$ is $|\psi_t|^2$ distributed (that is, $\scp{\psi_t}{P(\cdot)|\psi_t}$ distributed). 

The jump rate formula \eqref{Belljumprate} can be thought of as an analog of Bohm's equation of motion \eqref{Bohm} for jumps: for example, it involves quadratic expressions in $\psi$ in both the numerator and the denominator and leads to the equivariance of $|\psi|^2$. The point of the jump law is to set up a process $Q(t)$ once a Hilbert space $\Hilbert$, a state vector $\Psi\in\Hilbert$, a (reasonable) Hamiltonian $H$, a configuration space $\conf$, and configuration operators $P(dq)$ are given. Together with Bohm's equation of motion \eqref{Bohm}, the rate formula \eqref{Belljumprate} achieves this for Hamiltonians with ultraviolet cutoff, which brings us to the problem of ultraviolet divergence.

\subsection{An Ultraviolet Divergence Problem}

For the sake of concreteness of our discussion, consider a simplified, non-relativistic model QFT, in which $x$-particles can emit and absorb bosonic $y$-particles. Let us suppose that there is only 1 $x$-particle, and it is fixed at the origin, so $\Hilbert$ is the bosonic Fock space of the $y$-particles, and the configuration space is given by \eqref{confdef}. 

The naive, original expression for the Hamiltonian in the particle-position representation with creation and annihilation of $y$-particles at the origin $\vzero$ reads
\begin{align}
(H_\orig \psi)(\vy_1... \vy_n) &= 
-\frac{\hbar^2}{2m_y} \sum_{j=1}^n \nabla_{\vy_j}^2 \psi(\vy_1... \vy_n) \nonumber\\
&\quad +\: g \sqrt{n+1} \: \psi(\vy_1... \vy_n,\vzero)\nonumber\\
&\quad +\: \frac{g}{\sqrt{n}} \sum_{j=1}^n  \delta^3(\vy_j)\,\psi(\vy_1 ... \widehat{\vy_j} ... \vy_n)\,,\label{Horigdef}
\end{align}
where $g$ is a real coupling constant (the charge of the $x$-particle), and $\widehat{\vy_j}$ means that $\vy_j$ is omitted. Recall that $\psi$ is a function on $\cup_{n=0}^\infty \RRR^{3n}$, so $\psi(\vy_1... \vy_n)$ makes sense for any number $n$; note that $\psi(\vy_1... \vy_n,\vzero)$ refers to the $n+1$-sector of $\psi\in\Hilbert$ and $\psi(\vy_1 ... \widehat{\vy_j} ... \vy_n)$ to the $n-1$-sector. Roughly speaking, the middle line of \eqref{Horigdef} represents the annihilation of the $n+1$-st $y$-particle at the origin, while the last line represents the creation of a new $y$-particle at the origin, viz., with wave function $\delta^3$.

Unfortunately, the Hamiltonian \eqref{Horigdef} is ultraviolet (UV) divergent and thus mathematically ill defined. This means that the creation and annihilation terms in $H_\orig$, when expressed in the momentum representation, involve an integral over $\vk$ that diverges for large values of $|\vk|$. The root of the problem is that, according to the last line of \eqref{Horigdef}, the wave function of a newly created $y$-particle is a Dirac $\delta$ function, which has infinite energy and, what is worse, does not even lie in the Hilbert space (which contains only square-integrable functions). Many QFTs suffer from similar UV problems.

\begin{figure}[h]
\begin{center} 
\includegraphics[width=0.3\textwidth]{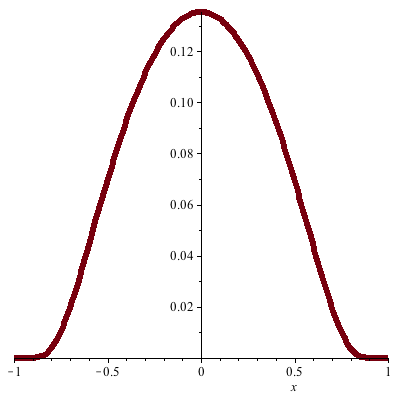}
\end{center}
 \caption{An example of a natural candidate for the cut-off function $\varphi(\cdot)$: a bump-shaped function that is a smooth and square-integrable approximation to a Dirac $\delta$ function and vanishes outside a small ball around the origin.}\label{cutoff}
\end{figure}

The UV problem can be circumvented by introducing an UV cut-off, i.e., by replacing the $\delta$ function by a square-integrable approximation $\varphi$ as in Figure~\ref{cutoff}.

\begin{figure}[h]
\begin{center} 
\includegraphics[width=0.6\textwidth]{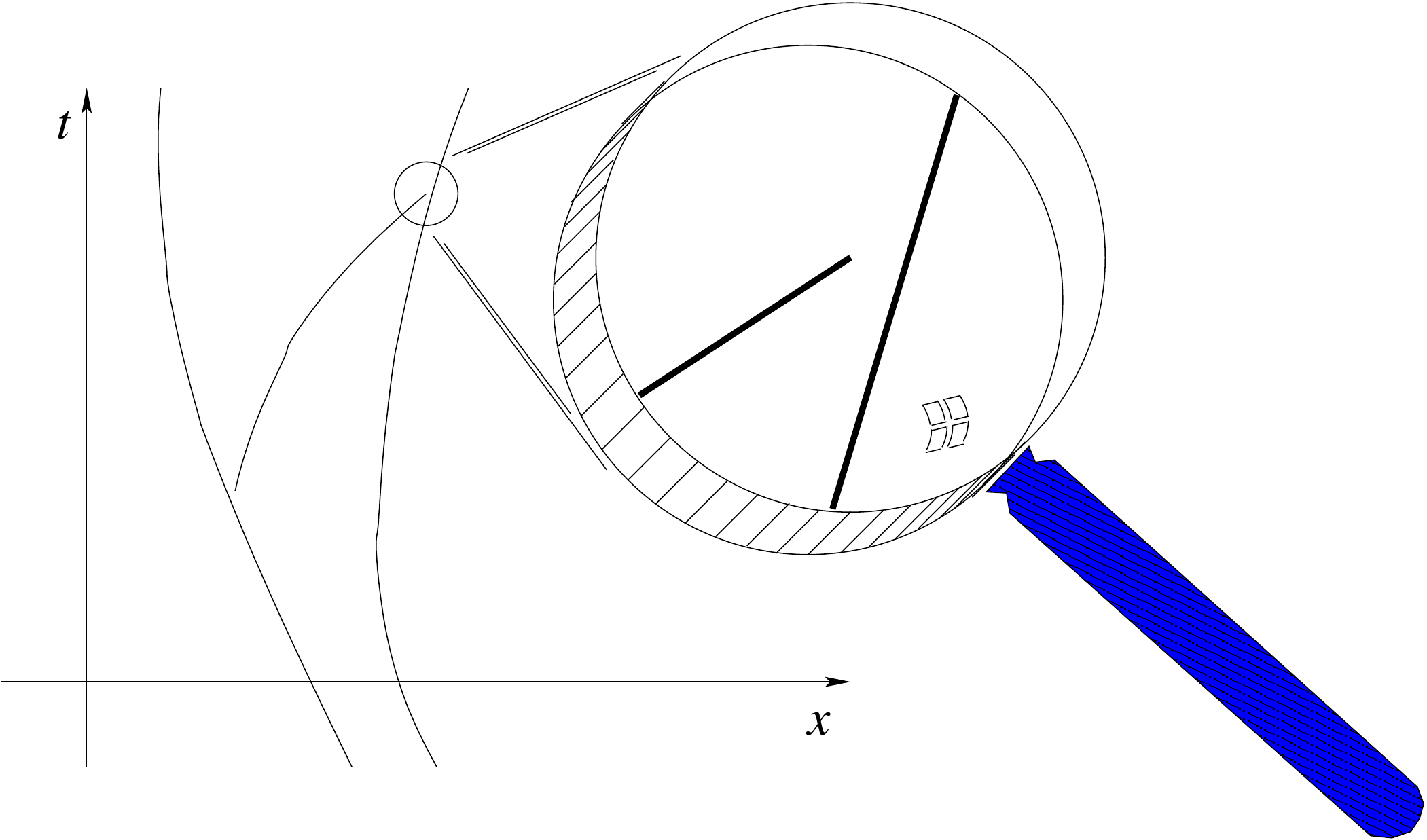}
\end{center}
 \caption{When using $H_\cutoff$, the emission and absorption of a $y$-particle happens, according to \eqref{Belljumprate}, not exactly at the location of an $x$-particle, but at a separation that can be as large as the radius of the support of $\varphi$. This does not happen with the alternative Hamiltonian defined by means of interior-boundary conditions.}\label{lookingglass}
\end{figure}

The cutoff corresponds to ``smearing out'' the $x$-particle
with ``charge distribution'' $\varphi(\cdot)$, and it leads to a well-defined Hamiltonian, given explicitly by
\begin{align}
(H_\cutoff \psi)(\vy_1 ... \vy_n) 
&= -\frac{\hbar^2}{2m_y} \sum_{j=1}^n \nabla_{\vy_j}^2 \psi(\vy_1 ... \vy_n) \nonumber\\
&\quad +\: g \sqrt{n+1} \int_{\RRR^3} \!\! d^3\vy\: \varphi^*(\vy)\: \psi\bigl(\vy_1 ... \vy_n,\vy \bigr) \nonumber\\
&\quad +\: \frac{g}{\sqrt{n}} \sum_{j=1}^n \varphi(\vy_j) \:\psi\bigl(\vy_1 ...  \widehat{\vy_j} ... \vy_n\bigr) \,. \label{Hcutoffdef}
\end{align}
However, there is no empirical evidence that electrons have a nonzero radius; it is therefore unknown which size or shape $\varphi$ should have; a cutoff tends to break Lorentz invariance; and, as another implausible consequence of the cutoff, emission and absorption occurs anywhere in the support of $\varphi$ around the $x$-particle, as depicted in Figure~\ref{lookingglass}.

\subsection{UV Problem Solved!}

Recent work \cite{TT15a,LSTT17,LS18,Lam18} has shown that this UV problem can be solved, at least in the non-relativistic case, by means of interior-boundary conditions (IBCs): they allow the rigorous definition of a Hamiltonian $H_\mathrm{IBC}$. In fact, for the specific Hamiltonian \eqref{Horigdef} with the $x$-particle fixed at the origin, it was known before \cite{Der03} that, for any sequence $\varphi_n \to \delta^3$, there exist constants $E_n\in \RRR$ such that $H_\cutoff-E_n$ possesses a limit $H_\infty$ as $n\to\infty$, called the renormalized Hamiltonian and independent of the choice of the sequence $\varphi_n$. It has been shown \cite{LSTT17} that $H_\infty$ coincides with $H_\mathrm{IBC}$ up to addition of a constant (i.e., of a multiple of the identity). However, for the case of moving $x$-particles in 3 space dimensions, it is not known how to obtain a renormalized Hamiltonian, and the IBC approach has provided for the first time a mathematically well defined Hamiltonian \cite{Lam18}.

\begin{figure}[h]
\begin{center} 
\includegraphics[width=0.4\textwidth]{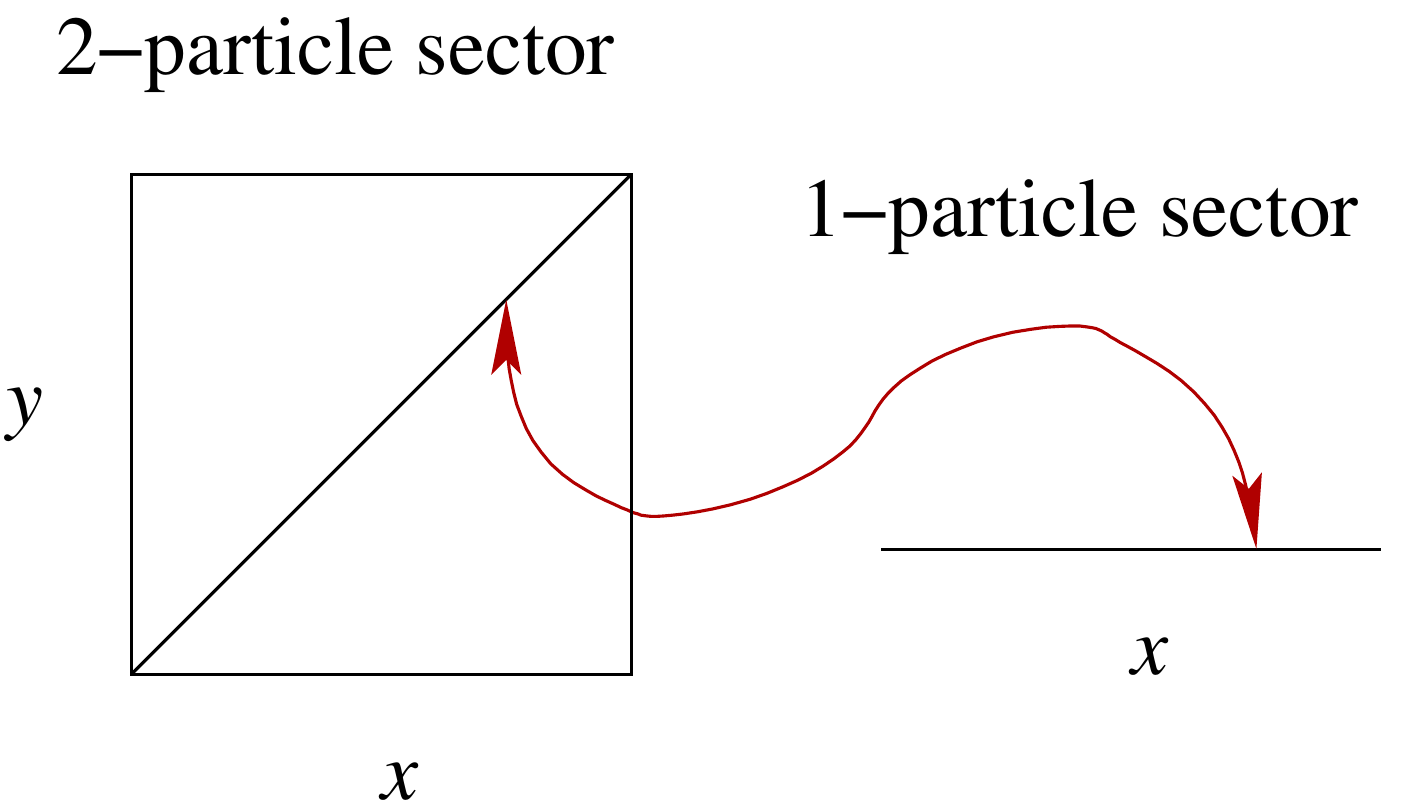} 
\end{center}
 \caption{An interior-boundary condition is a relation between the values of $\psi$ at two points: a point $q$ on the boundary (that is, where two particles collide, such as $(x,x)$ in the two-particle sector) and a point $q'$ in the interior of a lower sector (such as $x$).}\label{fig:ibc}
\end{figure}

Here is how this approach works \cite{TT15a,TT15b,Tum04,KS15}. An interior-boundary condition is a condition that links two configurations connected by the creation or annihilation of a particle, see Figure~\ref{fig:ibc}. Abstractly speaking, an IBC on a function $\psi$ on a domain $\Q$ with boundary $\partial\Q$ is a condition of the form
\be\label{ibc1}
\psi(q') = (\mathrm{const.}) \, \psi(q)\,,
\ee
where $q'$ is a boundary point and $q$ an interior point. In our case, the boundary configurations are those in which a $y$-particle meets an $x$-particle. In the case of moving $x$-particles, such configurations lie on diagonal surfaces in configuration space as depicted in Figure~\ref{fig:ibc}; in the case of a fixed $x$-particle at $\vzero$, they lie on the surfaces $\vy_k=\vzero$ for any $k=1,2,\ldots$. The corresponding interior configuration $q$ is the one with this $y$-particle removed, so $q$ lies one sector lower than $q'$. For example, with an $x$-particle at $\vzero$, the IBC is roughly of the form
\be\label{ibc2}
\psi(\vy_1... \vy_n, \vzero) = \frac{g\,m_y}{2\pi\hbar^2\sqrt{n+1}} \: \psi(\vy_1... \vy_n) \,.
\ee
In fact, the precise formula is yet a little different. That is because $|\psi|^2$ must diverge like $1/r^2$ as $r=|\vy|\to 0$ in order to guarantee a non-vanishing flux of probability into the origin; in fact, the relevant $\psi$s can be expanded in the form
\be\label{expansion}
\psi(\vy_1...\vy_n,\vy) = \alpha(\vy_1...\vy_n)\, r^{-1} + \beta(\vy_1...\vy_n) \, r^0 + o(r^0) 
\ee
($r=|\vy|$), and it is the leading coefficient $\alpha$ in this expansion
that should appear on the left-hand side of \eqref{ibc2}. Thus, the IBC reads
\be\label{ibc3}
\lim_{r\searrow 0} \, r\psi(\vy_1...\vy_n,r\vomega)
= \frac{g\,m_y}{2\pi\hbar^2\sqrt{n+1}} \: \psi(\vy_1...\vy_n)
\ee
for all unit vectors $\vomega\in\RRR^3$, $|\vomega|=1$. (The limit $r\searrow 0$ means $r\to 0$ with $r>0$.)

The expression for the corresponding Hamiltonian $H_{IBC}$ then reads, with $\SSS^2=\{\vomega\in\RRR^3: |\vomega|=1\}$ the unit sphere,
\begin{align}
(H_\mathrm{IBC}\psi)(\vy_1...\vy_n) 
&= -\frac{\hbar^2}{2m_y} \sum_{j=1}^n \nabla_{\vy_j}^2 \psi(\vy_1 ... \vy_n) \nonumber\\
&\quad + \: \frac{g\sqrt{n+1}}{4\pi}\int_{\SSS^2} d^2\vomega \, \lim_{r\searrow 0} \frac{\partial}{\partial r} \Bigl( r \psi(\vy_1...\vy_n,r\vomega) \Bigr) \nonumber\\[2mm]
&\quad +\: \frac{g}{\sqrt{n}} \sum_{j=1}^n  \delta^3(\vy_j)\,\psi(\vy_1...\widehat{\vy_j}...\vy_n) \,.\label{Hibc}
\end{align}
The term in the last line, involving the problematical $\delta$ function, actually gets canceled by the term created when the Laplacian gets applied to the $\alpha r^{-1}$ term in \eqref{expansion}, which contributes a $\delta$ function; the constant prefactor in the IBC \eqref{ibc2} or \eqref{ibc3} is dictated by the goal of this cancellation. The middle line extracts the next-to-leading coefficient $\beta$ of \eqref{expansion} from $\psi$ in the last variable $\vy_{n+1}$. (As a consequence of the expansion \eqref{expansion}, which is valid for $\psi$ in the domain of $H_\mathrm{IBC}$, the integrand is independent of $\vomega$, so that it is actually unnecessary to average over $\vomega$.)

Here is the rigorous result about $H_\mathrm{IBC}$:

\paragraph{Theorem \cite{LSTT17}}
{\it On a suitable dense domain $\domain_\mathrm{IBC}$ of $\psi$s in $\Hilbert$ of the form \eqref{expansion} satisfying the IBC \eqref{ibc3}, $H_\mathrm{IBC}$ is well defined, self-adjoint, and positive. In particular, there is no UV divergence.}

\bigskip

Historically, IBCs were invented several times for various purposes \cite{Mosh51a,Mosh51b,Tho84,Yaf92}, but only recently considered for the UV problem \cite{TT15a,TT15b}. Rigorous results about existence and self-adjointness of the Hamiltonian were proved in \cite{Lam18} for moving $x$-particles in 3 dimensions, in \cite{LS18} for moving $x$-particles in 2 dimensions, and also in \cite{LS18} for the Nelson model \cite{Nel64} in 3 dimensions.

\subsection{Particle Trajectories}

Also to $H_\mathrm{IBC}$ there is associated a jump process in $\Q$ analogous to Bell's that is $|\psi_t|^2$ distributed at every time $t$ \cite{bohmibc}. In this process, the world lines of $y$-particles begin and end on those of the $x$-particles (like in Figure~\ref{crletf1}(a) and unlike in Figure~\ref{lookingglass}). We conjecture that this process is the limit of the continuum Bell process governed by \eqref{Belljumprate} as $\varphi\to\delta^3$.

Since the Hamiltonian is no longer of the form $H_0+H_I$ (particularly as the functions in the domain of $H_0$ do not satisfy the boundary condition), the jump rate formula \eqref{Belljumprate} does not immediately apply. Nevertheless, the process can be defined as follows \cite{bohmibc}. Between the jumps, the configuration follows Bohm's equation of motion in $\Q^{(n)}=\RRR^{3n}$. Every jump is either an absorption (to the next lower sector) or an emission (to the next higher sector). The absorption events are deterministic and occur when $Q(t)\in\Q^{(n)}$ reaches $\vy_j=\vzero$ for any $j=1...n$; in that moment, the configuration jumps to $(\vy_1...\widehat{\vy_j}...\vy_n)\in\Q^{(n-1)}$. The emission of a new $y$-particle at $\vzero\in\RRR^3$ occurs at a random time $t$ in a random direction $\vomega$ (there is one trajectory starting there in each direction $\vomega$) with a rate dictated by time reversal invariance, the Markov property, and the wish for equivariance \cite{Tum04,bohmibc}:
If $Q(t)=y=(\vy_1...\vy_n)\in\Q^{(n)}$, then with jump rate
\be\label{jumprate2}
\sigma^\psi(y \to y \times 0d^2 \vomega) 
=  \lim_{r\to 0}\frac{\max\bigl\{0,\tfrac{\hbar}{m} \, \Im\bigl[r^2 \psi(y,r\vomega)^* \, \partial_r \psi(y,r\vomega) \bigr]\bigr\}}{|\psi(y)|^2} d^2\vomega
\ee
it jumps to the solution of Bohm's equation of motion in $\Q^{(n+1)}$ beginning at
\be
(\vy_1, \ldots, \vy_{j-1}, 0\vomega, \vy_j,\ldots, \vy_n)
\ee
with $1\leq j\leq n+1$. That is, the newly created $y$-particle at the origin gets inserted at the $j$-th position, where $j$ is chosen uniformly random ($\psi$ is symmetric against permutation), and starts moving in direction $\vomega$. By virtue of \eqref{expansion}, the right-hand side of \eqref{jumprate2} is actually independent of $\vomega$, so $\vomega$ is random with uniform distribution.

\section{Extension of Bohmian Mechanics to Relativistic Space-Time}
\label{sec:rela}

\subsection{The Time Foliation}

A \emph{foliation} is a slicing of space-time into hypersurfaces, that is, a family of non-overlapping hypersurfaces whose union is space-time. We will consider the possibility that there is a \emph{preferred} foliation of space-time into spacelike hypersurfaces (``time foliation'' $\foliation$), that is, that one foliation $\foliation$ plays a special dynamical role in nature, essentially defining a kind of simultaneity at a distance. If the existence of a time foliation is granted, then there is a simple, convincing analog of Bohmian mechanics, BM$_{\foliation}$. For a single particle, a time foliation is unnecessary, as Bohm found already in 1953 \cite{Bohm53}.
Bohm and Hiley \cite{BH} introduced the equation of motion of BM$_{\foliation}$ for flat foliations (i.e., parallel hyperplanes, i.e., Lorentz frames), D\"urr et al.~\cite{HBD} for curved foliations, and I contributed \cite{Tum01} a proof of equivariance for curved space-time. The surfaces belonging to $\foliation$ will be called the \emph{time leaves}.

\begin{figure}[h]
\begin{center} 
\includegraphics[width=0.4\textwidth]{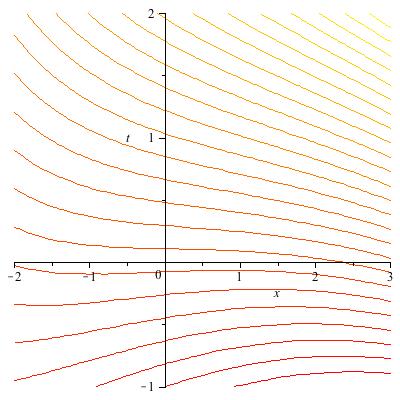}
\end{center}
 \caption{Example of a spacelike foliation (i.e., slicing into spacelike hypersurfaces) of Minkowski space-time in $1+1$ dimensions}\label{foliation}
\end{figure}

Without a time foliation (i.e., a preferred foliation), no version of Bohmian mechanics is known that would make predictions anywhere near quantum mechanics, and I have no hope that such a version can be found in the future. 

Sutherland \cite{Suth08,Suth17} has made an attempt towards such a version; he has proposed a Bohm-like equation of motion without a time foliation but involving retrocausation. While one may have reservations about retrocausation, it would be of interest to know whether such a theory can be made to work. At the present stage, Sutherland has formulated a proposal for trajectories of non-interacting particles between measurements at times $t_i$ and $t_f$; for an assessment, one would need to formulate a proposal that can be applied to the universe as a whole and that can also treat measurements as just particular instances of motion and interaction of particles. I have considered a natural extension of Sutherland's equations to a universe with interaction and concluded that measurement outcomes, if their records get erased before the final time of the universe, may have a probability distribution that deviates very much from the one predicted by quantum mechanics and BM$_\foliation$. So one would have to come up with a better proposal for an interacting version.

Let me return to BM$_\foliation$. To grant a time foliation seems against the spirit of relativity. But it is a real possibility that our world is like that. It does not mean relativity would be irrelevant: After all, there is still a metric $g_{\mu\nu}$; the free Hamiltonian is still the Dirac operator (or whichever relativistic operator is appropriate); formulas are still expressed with 4-vector indices ($j^{\mu}$ etc.); the statistics of experimental outcomes are independent of $\foliation$ (see below); and superluminal signaling is impossible in BM$_\foliation$. On the other hand, there exists also the vector $n_\mu$ normal to the time foliation, and the hypothesis of a time foliation provides a simple and straightforward explanation of the non-locality required by Bell's theorem.

A preferred foliation may be provided anyhow by the metric: If we take space-time to be curved and have a big bang singularity (which seems realistic), then the simplest choice of $\foliation$ consists of the level sets of the real-valued function $T$ on space-time such that $T(x)$ is the timelike distance of $x$ from the big bang; e.g., $T$(here-now) = 13.7 billion years (if what we call the big bang did involve a singularity).

Alternatively, $\foliation$ might be defined in terms of the quantum state vector $\psi$, $\foliation=\foliation(\psi)$ \cite{DGNSZ14}, or $\foliation$ might be determined by an evolution law (possibly involving $\psi$) from an initial time leaf. 

Let us turn to the definition of the trajectories.

\subsection{The Single-Particle Case}

I begin with the simplest case, that of a single particle \cite{Bohm53}, which does not involve the time foliation $\foliation$. Let $\psi:\RRR^4\to\CCC^4$ be a solution of the Dirac equation
\be\label{Dirac}
i\hbar\gamma^\mu \partial_\mu \psi = m\psi \,.
\ee
The vector field
\be\label{jmu}
j^\mu = \overline{\psi} \gamma^\mu \psi
\ee
is called the probability current 4-vector field. It is formed in a covariant way (since $\psi\mapsto \overline\psi = \psi^\dagger \gamma^0$ is a covariant operation, whereas $\psi\mapsto \psi^\dagger$ is not); $j^\mu$ is real, future timelike-or-lightlike, and divergence free, $\partial_\mu j^\mu=0$.

The Bohmian trajectories are the integral curves of the vector field $j^\mu$; put differently, the equation of motion reads
\be\label{BD}
\frac{dQ^\mu}{d\tau} \propto j^\mu(Q^\nu(\tau)) \,,
\ee
where $\tau$ can be proper time or, in fact, any curve parameter, and $\propto$ means ``is proportional to.'' In fact, it suffices to prescribe $dQ^\mu/d\tau$ only up to scalar factors (and to allow any curve parameter) because that fixes the tangent (i.e., the direction) of the world line in space-time.

It then follows that the possible world lines are timelike-or-lightlike curves. On any spacelike (Cauchy) hypersurface $\Sigma_0$, we can choose an initial condition $Q^\mu(\tau=0) \in \Sigma_0$, and a unique solution curve $Q^\mu(\tau)$ exists for all times (except, technically speaking, for a set of measure zero of initial conditions) \cite{TT05}. Equivariance holds in the following sense: On a spacelike (Cauchy) hypersurface $\Sigma$, the appropriate interpretation of ``$|\psi|^2$ distribution'' is the distribution whose density relative to the 3-volume $d^3x$ defined by the 3-metric on $\Sigma$ is $j^\mu n_\mu = \overline{\psi}  n\hspace{-2mm}/ \psi$ with $n_\mu(x)$ the future unit normal vector to $\Sigma$ at $x\in\Sigma$ and $n\hspace{-2mm}/ = n_\mu \gamma^\mu$. If the initial condition $Q^\mu(\tau=0)$ is random with distribution $|\psi_{\Sigma_0}|^2$ then on every other $\Sigma$, the intersection point of the world line with $\Sigma$ is random with distribution $|\psi_{\Sigma}|^2$. The evolution of $\psi$ from $\Sigma_0$ to $\Sigma$ is unitary.

All I said remains true when an external electromagnetic field is added to the Dirac equation, or when we consider a curved space-time.

\subsection{Law of Motion for Many Particles}

\begin{figure}[h]
\begin{center} 
\includegraphics[width=0.4\textwidth]{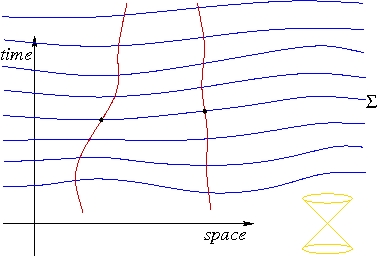}
\end{center}
 \caption{The equation of motion of BM$_\foliation$ specifies the tangent direction of a world line by means of the wave function evaluated at the configuration where all world lines intersect the same time leaf $\Sigma$.}\label{foliation2}
\end{figure}

Here is the definition of BM$_\foliation$ \cite{HBD}.
Consider $N$ particles. Suppose that,
for every $\Sigma\in\foliation$, we have a wave function $\psi_\Sigma$ on $\Sigma^N$. (We will discuss in the next section how to obtain $\psi_\Sigma$ from multi-time wave functions.) For $N$ timelike-or-lightlike world lines $Q_1\ldots Q_N$, the configuration on $\Sigma$ consists of the intersection point of each world line with $\Sigma$,
\be
Q(\Sigma) = (Q_1\cap\Sigma,\ldots,Q_N\cap\Sigma)
\ee
The equation of motion is of the form (see Figure~\ref{foliation2})
\be\label{expression}
\frac{dQ_k^\mu}{d\tau}\propto \mathrm{expression}\Bigl[\psi\bigl(Q(\Sigma)\bigr)\Bigr] \,.
\ee

Specifically, for $N$ Dirac particles, the wave function is of the form $\psi_\Sigma: \Sigma^N \to (\CCC^4)^{\otimes N}$ for every $\Sigma\in\foliation$, and the equation of motion reads
\be\label{HBD}
\frac{dQ^\mu_k}{d\tau} \propto
j_k^\mu(Q(\Sigma)),
\ee
where
\be
j^{\mu_1\ldots \mu_N}(x_1...x_N) = \overline{\psi} (x_1...x_N)[\gamma^{\mu_1}\otimes \cdots \otimes \gamma^{\mu_N}] \psi(x_1...x_N)\,,
\ee
\be
j_k^{\mu_k}(x_1... x_N) = j^{\mu_1\ldots \mu_N}(x_1...x_N)
\, n_{\mu_1}(x_1) \cdots (k \text{-th omitted}) \cdots n_{\mu_N}(x_N)\,,
\ee
and $n_\mu(x)$ is the future unit normal vector to $\Sigma$ at $x\in\Sigma$.

The appropriate version of the $|\psi|^2$ distribution (which we will simply call $|\psi|^2$) is the one with density 
\be
\rho(x_1...x_N)=j^\mu_k(x_1...x_N) \, n_\mu(x_k)= \overline\psi [n\hspace{-2mm}/(x_1)\otimes \cdots \otimes n\hspace{-2mm}/(x_N)]\psi
\ee
relative to the volume $d^3x_1 \cdots d^3x_N$ defined by the metric $g_{\mu\nu}$ on $\Sigma$. (Actually, $\rho$ is literally $|\psi|^2$ if for each $x_j$ we use the Lorentz frame tangent to $\Sigma$ at $x_j$.) It can be shown \cite{HBD,Tum01} that the $|\psi|^2$ distribution is equivariant, more precisely: If the initial configuration is $|\psi|^2$-distributed, then the configuration $Q(\Sigma)$ is $|\psi_\Sigma|^2$-distributed \emph{on every $\Sigma\in\foliation$}. Moreover:

\paragraph{Theorem \cite{LT17}}
{\it If detectors are placed along any spacelike surface $\Sigma$ (and if some reasonable assumptions about the evolution of $\psi_\Sigma$ are satisfied), then the joint distribution of detection events is $|\psi_\Sigma|^2$.}

\bigskip

That is, while undetected configurations $Q(\Sigma')$ may fail to be $|\psi_{\Sigma'}|^2$ distributed if $\Sigma'$ is not a time leaf, the detected configuration is $|\psi_\Sigma|^2$-distributed on \emph{every} spacelike $\Sigma$. As a consequence, $\foliation$ is invisible, i.e., experimental results reveal no information about $\foliation$. In fact, all empirical predictions of BM$_\foliation$ agree with the standard quantum formalism (and the empirical facts). 

BM$_\foliation$ is a very robust theory, as it works for arbitrary foliation $\foliation$; it works even if the time leaves have kinks \cite{ST14} (a case in which $\foliation$ violates a condition in the mathematicians' definition of ``foliation''); it works even if the leaves of $\foliation$ overlap \cite{ST15}; it can be combined with the stochastic jumps for particle creation; it works also in curved space-time \cite{Tum01}; and it still works if space-time has singularities \cite{Tum10}.

\subsection{Multi-Time Wave Functions}

A multi-time wave function $\phi(t_1,\vx_1,\ldots,t_N,\vx_N)$ \cite{dirac:1932,dfp:1932,bloch:1934,LPT:2017}
is a natural relativistic generalization of the $N$-particle wave function $\psi(t,\vx_1,\ldots,\vx_N)$ of non-relativistic quantum mechanics: It is a function of $N$ space-time points, and thus of $N$ time variables. It is usually defined only on the set $\sS$ of spacelike configurations, i.e., of those $N$-tuples $(x_1 \ldots x_N) \in \RRR^{4N}$ of space-time points $x_j=(t_j,\vx_j)\in\RRR^4$ for which any two $x_j,x_k$ are spacelike separated or identical. 
$\phi$ is the covariant particle-position representation of the state vector.
The usual (single-time) wave function $\psi$ is contained in $\phi$ by setting all time variables equal,
\be\label{psiphi}
\psi(t,\vx_1,\ldots,\vx_N) = \phi(t,\vx_1,\ldots,t,\vx_N) \,.
\ee
More generally, we can obtain for every spacelike hypersurface $\Sigma$ a wave function $\psi_\Sigma$ on $\Sigma^N$ by simply setting
\be
\psi_\Sigma(x_1,\ldots,x_N) = \phi(x_1,\ldots,x_N)
\ee
for all $x_1,\ldots,x_N \in \Sigma$.
This is the $\psi_\Sigma$ that goes into \eqref{expression}, \eqref{HBD}, and the theorem from \cite{LT17} reported in the previous subsection. Thus, the theorem is really a theorem about multi-time wave functions. Since $\psi_\Sigma$ is closely related to the Tomonaga-Schwinger \cite{tomonaga:1946,schwinger:1948} wave function, so is $\phi$; at the same time, $\phi$ is a simpler kind of mathematical object, as it is a function of only finitely many variables (at least locally, when we consider Fock space).

The obvious choice (though not the only possible one \cite{lienert:2018}) of time evolution equations for $\phi$ is to introduce an equation for each time variable, 
\be\label{multiSchr}
i\hbar\frac{\partial \phi}{\partial t_j} = H_j \phi \quad \forall j=1\ldots N.
\ee
It follows that the single-time wave function $\psi$ as in \eqref{psiphi} will evolve according to the usual kind of Schr\"odinger equation
\be
i\hbar\frac{\partial\psi}{\partial t}=H\psi
\ee
if and only if
\be
\sum_{j=1}^N H_j = H
\ee
at equal times, a relation relevant to guessing suitable multi-time Schr\"odinger equations \eqref{multiSchr}.

A big difference between multi-time and single-time Schr\"odinger equations is that for \eqref{multiSchr} to possess solutions for all initial conditions at $0=t_1=t_2=\ldots = t_N$, the partial Hamiltonians $H_j$ must satisfy a \emph{consistency condition} \cite{dfp:1932,bloch:1934,pt:2013a}
\be
\biggl[i\hbar\frac{\partial}{\partial t_j} - H_j, i\hbar\frac{\partial}{\partial t_k} -H_k\biggr] = 0
\quad \forall j\neq k \,.
\ee
If the $H_j$ are time-independent, then the condition reduces to $[H_j,H_k]=0$. These conditions are trivially satisfied for non-interacting particles \cite{schweber:1961}, but to implement interaction is a challenge; for example, interaction potentials violate consistency \cite{pt:2013a,ND:2016}. However, it has been shown that interaction can be consistently implemented \cite{DV85}, in particular in the form of zero-range interactions (``$\delta$ potentials'') \cite{lienert:2015a,LN:2015} and of interaction through emission and absorption of bosons \cite{pt:2013c,pt:2013d}.

The upshot is that the evolution of the wave function can be defined in a covariant way without using the time foliation $\foliation$, which then needs to be introduced for the trajectories. The evolution of the wave function can directly be formulated in the particle-position representation, in fact with rather simple equations \cite{pt:2013c,lienert:2018}.

\section{Outlook and Concluding Remarks}

Those who regard a theory with a preferred foliation as unacceptable may want to consider relativistic collapse theories instead \cite{Tum06,BDGGTZ}, which do not need a preferred foliation. I believe, however, that we should take the possibility of a preferred foliation (depending perhaps on the space-time metric and/or the wave function) seriously. Then BM$_\foliation$ seems like the most plausible ontological theory of quantum mechanics in relativistic space-time, and I regard it as a fully satisfactory extension of Bohmian mechanics to relativistic space-time. Particle creation and annihilation can be incorporated into it in the same way as described in Section~\ref{sec:crea} for the non-relativistic case.

A goal for the future would be to formulate a version of quantum electrodynamics (QED) with particle trajectories. The particle-position representation of the quantum state in QED was formulated already by Landau and Peierls \cite{lp:1930} in 1930, and it lends itself nicely to a multi-time formulation. So what are the obstacles? The main obstacle is that defining Bohmian trajectories for a photon requires defining the probability current $j^\mu$, so we would need a formula for photons analogous to $j^\mu = \overline{\psi}\gamma^\mu \psi$ for Dirac wave functions, but such a formula is not known to date except for plane waves (for which it is $j^\mu = |c|^2 k^\mu/\hbar$ whenever the energy-momentum tensor is $T^{\mu\nu}= |c|^2 k^\mu k^\nu$). Of course, this problem concerns not only the Bohmian approach but every approach to QED, but it is of particular importance in the Bohmian framework. Oppenheimer \cite{Opp31} argued in 1931 that $j^\mu$ does not exist for photons; while his argument is not completely compelling, it is by itself quite reasonable. However, since we can measure probability distributions of photons in photon counters and interference experiments, I have trouble imagining how $j^\mu$ could fail to exist for photons. So, I tend to suspect that there is a formula for $j^\mu$ which we have not found yet. 

Another problem for future research is whether the technique of interior-boundary conditions can be applied to relativistic Hamiltonians. A further problem is how to deal in the Bohmian framework with positrons, the Dirac sea, and states of negative energy. Some authors \cite{CS07,DEO17} have proposed to take the Dirac sea literally as an infinity (or at least a very large number) of Bohmian particles. I am inclined to take positrons literally as Bohmian particles, but various questions about this approach remain open.

Let me conclude. While standard quantum mechanics is often unclear, standard quantum field theory is often even less clear. But the developments I have described provide reasons for optimism that a clear version of serious QFTs (such as QED) can be obtained, and the Bohmian approach of using particle trajectories is in my opinion the most promising candidate for getting there. A fully satisfactory formulation of non-relativistic quantum mechanics is provided by Bohmian mechanics, and I believe that we should try hard to reach a clear formulation of QED as well. Some of the difficulties of QED are of a mathematical nature (such as the precise definition of the time evolution of the quantum state), others of an ontological nature (what is actually there), and yet others of an operational nature (such as how to compute the position probability distribution of photons for arbitrary states). Some of the difficulties can often be circumvented or ignored, while the Bohmian approach forces us to face them. I think that is ultimately an advantage.

\end{document}